\documentclass[a4paper,10pt]{article}
\usepackage[utf8]{inputenc}

\title{ Holographic Cavalieri Principle as a Universal relation between Holographic Complexity and Holographic Entanglement Entropy}
\author{Davood  Momeni$^{1}$, Mir Faizal$^{2, 3}$, Ratbay Myrzakulov$^4$\\\\$^1$ Department of Physics, College of Science, Sultan Qaboos University,\\
P.O. Box 36, P.C. 123, Al-Khowd, Muscat, Sultanate of Oman
\\$^2$ University of British
Columbia - Okanagan \\
  Kelowna,  British Columbia V1V 1V7, Canada
\\$^3$ Department of Physics and Astronomy, University of Lethbridge, \\
Lethbridge, Alberta, T1K 3M4, Canada
\\$^4$  Eurasian International Center for Theoretical Physics \\
and Department of General Theoretical Physics, \\
Eurasian National University, Astana 010008, Kazakhstan
}
\date{}
\begin{document}

\maketitle

\begin{abstract}
In this paper, we will propose a universal relation between the holographic
  complexity (dual to a volume in AdS) and  the holographic entanglement entropy (dual to an area in AdS).
We will   explicitly demonstrate that our conjuncture hold for all a metric 
asymptotic to AdS$_3$, and then argue   that such a relation should hold in general due 
to the  AdS version of the   Cavalieri principle. We will   demonstrate that it holds for 
Janus solution, 
which have been recently been obtained in type IIB string theory. 
We will also show that this conjecture holds for a circular disk.  
This conjecture will be used to show that the proposal that the complexity equals action, and the proposal that the complexity 
equal volume can represent the same physics. Thus, using this conjecture, we will show that the black holes are fastest computers, 
using the proposal that complexity equals volume. 
\end{abstract}

\section{Introduction}
Various studies done in different areas of  physics have indicated that  the laws of physics can be  represented   in terms of 
the ability of an observer to  process relevant information \cite{info, info2}.  
Entropy  measures the amount of information that is lost in a process, and hence, it 
is thought to be one of the most important quantities associated with any such information theoretical process. 
The entropy has been used to model physical phenomena from condensed matter physics to gravitational physics. Even the geometry of spacetime  can be 
viewed as an emergent structure, which emerges  due to 
 an information theoretical process.
This is because  in the Jacobson formalism, the Einstein equation can be derived from thermodynamics by assuming a certain scaling  behavior 
of the  entropy    \cite{z12j, jz12}. This scaling behavior of entropy is that the maximum entropy of a region of space scales with its  
area, and this observation has been   obtained using the physics  of black holes.  This observation has also led to the development of the 
holographic principle \cite{1, 2}, and the AdS/CFT correspondence is one of the most
important realizations of the holographic principle  \cite{M:1997}. 

The black hole information paradox occurs due to the 
observation that black holes are maximum entropy objects and they evaporate due to Hawking radiation. 
It is interesting to note that  quantum entanglement  has   been used to discuss the   microscopic nature 
of black hole entropy, with the hope that it may resolve the black hole information paradox \cite{4,5}.
The AdS/CFT correspondence makes it possible to quantify   quantum 
entanglement in terms of the holographic
entanglement entropy \cite{6,RT,6a}. The  entanglement entropy has been used in various branches of physics from quantum computing to 
condensed matter physics, and AdS/CFT correspondence make it possible to calculate it holographically. The holographic entanglement entropy 
 of a CFT   is dual to the area of a minimal surface defined in the bulk of an asymptotically AdS spacetime. So, 
for  subsystem $A$ with its complement, it is possible to write an expression for the holographic entanglement entropy as
\begin{equation} \label{HEE}
S_{A}=\frac{\mathcal{A}(\gamma _{A})}{4G_{d+1}}
\end{equation}
where $G$  is the gravitational constant in the AdS spacetime,  $\gamma_{A}$ is the $(d-1)$-minimal surface extended 
into the AdS bulk with the boundary $\partial A$, and $\mathcal{A}(\gamma _{A})$ is the area of this minimal surface. 
It may be noted that usually there are UV divergence in holographic entanglement entropy, and so we need to use 
a regularization method to remove these divergences. Thus, for a deformed geometry, we define the area in this paper as,
\begin{eqnarray}
 \mathcal{A}(\gamma_A) = \mathcal{A}_{D}(\gamma_A) - \mathcal{A}_{AdS}(\gamma_A), 
\end{eqnarray}
where $ \mathcal{A}_{D}(\gamma_A) $ is the defined in deformed geometry (for example the geometry of a black hole), and $ \mathcal{A}_{AdS}(\gamma_A)$
is defined in the background $AdS$ spacetime. Thus, we define the holographic entanglement entropy for a deformed geometry by subtracting the 
contribution coming from the background $AdS$ spacetime. This removes the divergent part and we are only left with a finite part. 
We will use this finite part in this paper, and call it the holographic entanglement entropy. 

However, the recent studies have indicated that it is not enough to know what part of the information can be obtained 
by an observer from a system, but it is also important to know how difficult is it to obtain that information. 
As the entropy quantifies the abstract notion of the loss of information, the complexity quantifies the abstract notion 
of the difficulty to obtain the information (even if it is present in the system). 
The complexity (like entropy) has been used to study physical systems from black holes to condensed matter physics, and even 
quantum computing. In fact, 
recently it has been proposed that the information may not be ideally lost in a black hole, but it may be 
lost for all practice purposes as it would be impossible to reconstruct it from the Hawking radiation \cite{hawk}. 
As complexity has only been recently  used to study various physical systems, there are different proposals to define the 
complexity for a CFT. However, recently 
motivated  by  holographic entanglement entropy,  
 holographic complexity has been holographically defined as a quantity dual to   a volume of  codimension one time 
slice in anti-de Sitter (AdS) \cite{Susskind:2014rva1}-\cite{Couch:2016exn}. 
Furthermore, it is possible to use a subsystem $A$ with its complement, and define this volume as  $V = V(\gamma_A)$, 
i.e., the volume enclosed by the same  minimal surface which was used to calculate the holographic 
entanglement entropy \cite{Alishahiha:2015rta}, 
\begin{equation}\label{HC}
\mathcal{C}_A= \frac{V(\gamma_A) }{8\pi R G_{d+1}},
\end{equation}
where $R$ and $V(\gamma_A)$ are the radius of the curvature and the volume in the AdS bulk.  
So, we will use this definition of the  holographic complexity, and investigate a relation between this definition of holographic complexity
 and holographic entanglement entropy. It may be noted that just 
 like the minimal area, this volume also contains UV divergences, and  so we need to 
regularize  this volume. So,for a deformed geometry, we define the volume as 
\begin{eqnarray}
V(\gamma_A) = V_{D}(\gamma_A) - V_{AdS}(\gamma_A), 
\end{eqnarray}
where $ V_{D}(\gamma_A) $ is the volume  in deformed geometry,  and $V_{AdS}(\gamma_A)$
is the volume in the background $AdS$ spacetime. So,we regularize the volume in a deformed geometry by subtracting the  
contribution coming from the background $AdS$ spacetime. This again removes the divergent part and we are again  left with a finite part. 
In this paper, we will use this finite part of the holographic complexity.

Now as both the holographic complexity and holographic entanglement entropy are calculated using the same   minimal surface, we expect that 
a  universal relation to exist between them due to an AdS version of Cavalieri principle. In this letter, we will explicitly demonstrate this to be the 
case, and also  find the explicit form of this universal relation. This can be used as a new  holographic dictionary to calculate the holographic 
complexity from holographic entanglement entropy  for different asymptotic AdS spacetimes. As both holographic entanglement entropy and complexity 
are used in various different branches of physics ranging from black hole physics to condensed matter physics, this general conjecture can have 
a lot of applications in those branches. This is because it is easier to calculate the entanglement entropy than complexity for various complex systems, 
and if this conjecture holds in general, this can be used as a holographic dictionary to obtain such quantities. Furthermore, as it is difficult to define 
complexity for the boundary theory, and a precise definition for complexity does not exist for the boundary theory, we will use this relation 
to obtain a definition for complexity of the boundary theory. This is because the complexity will be defined in terms of quantities whose boundary dual is 
well understood.

\section{Excited States in Bulk Geometry}  
In this section, we will motivate a universal relation between the holographic complexity and holographic entanglement entropy. 
 Now we consider  an excited state
in a $d+1$ dimensional conformal field theory (CFT), and  assume it to be 
almost static and translational invariant. We want to analyse its gravity dual, so we write the metric on  $AdS_{d+2}$
as \cite{Bhattacharya:2012mi} 
\begin{eqnarray}
ds^2=\frac{R^2}{z^2}\left[-f(z)dt^2+\frac{dz^2}{h(z)}+\sum^{d}_{i=1}(dx_i)^2\right].\label{pads}
\end{eqnarray}
 Near the boundary $z\to 0$, we can assume $h(z)\simeq
f(z)\simeq 1-\big(\frac{z}{z_0}\big)^{d+1}$, where $z_0$ is constant.

Let us consider an entangling region (subsystem $A$) in the shape of  a strip defined by $0<x_1<l,\ \
\ -L/2<x_{2,3,...,d}<L/2$, where $L$ is taken to be infinite. So, we
can parameterize the minimal surface $\gamma_A$ by $x_1=x(z)$, and write its area as 
\begin{eqnarray}
\mathcal{A}(\gamma _{A})=2L^{d-1}R^{d}\int\frac{dz}{z^{d}}\sqrt{\frac{1}{h(z)}+x'_1(z)^2}
\label{Area}.
\end{eqnarray}
where the derivative with respect to the $z$ is denotes by $'$. 
 We can determine the shape
of $x(z)$ by minimizing this area function.  We note that the Lagrangian in Eq. (\ref{Area}) is independent of $x_1(z)$, 
and so the first integral  associated with Euler-Lagrange equation can be expressed as 
\begin{eqnarray}
&& x'_1(z)=\frac{1}{h(z)}\frac{(\frac{z}{z^{*}})^{d+1}}{\sqrt{1-(\frac{z}{z^{*}})^{2(d+1)}}}\label{x1}.
\end{eqnarray}
here we have assumed that $x'_1(z^{*})=\infty$. 
The total entangled length $l$, the entanglement area functional $\mathcal{A}(\gamma _{A})$ and volume of
codimension one time slice $V(\gamma_A)$ of the metric (\ref{pads}) are given by 
\begin{eqnarray}
&&l=2\int_{\epsilon}^{z^{*}}dz\big(\frac{z}{z^*}
\big)^d\sqrt{\frac{1}{h(z)(1-\big(\frac{z}{z^*}
\big)^{2d})}} \label{l}
\\&& \mathcal{A}(\gamma _{A})=\frac{L^{d-1}R^{d}}{(z^{*})^d}
\int_{\frac{\epsilon}{z^{*}}}^{1}\frac{d\xi}{\xi^{d+1}}\sqrt{\frac{1}{h(\xi)(1-\xi^{2(d+1)})}}\label{A}
\\&&
V(\gamma_A)=\frac{L^{d-1}R^{d}}{(z^{*})^{d-1}}\int_{\frac{\epsilon}{z^{*}}}^{1}\frac{x_1(\xi)d\xi}{\xi^{d+1}\sqrt{h(\xi)}}\label{V}.
\end{eqnarray}

It may be noted that to evaluate integral involving $l$, we   have  defined     $\xi=\frac{z}{z^{*}}$,    $h(z)\sim 1-\xi^{d+1}a^{d+1}$, and 
$a=\frac{z^{*}}{z_0}$. 
Now in general,  $\epsilon\to 0$, thus it is adequate to just evaluate integral for dominant terms
in the region $\xi\ll 1$. Using a series expansion in integrand, which is valid for all $\frac{\epsilon}{z^{*}}<\xi\ll1$, we obtain, 
\begin{eqnarray}
l\approx\frac{2z^{*}}{d+1}\big(1-(\frac{\epsilon}{z^{*}})^{d+1})+\frac{z^{*} a^{d+1}}{2(d+1)}\big(1-(\frac{\epsilon}{z^{*}})^{2(d+1)}).
\end{eqnarray}
In limit $\epsilon\to 0$, we obtain finite and regular length, 
\begin{eqnarray}
l\approx\frac{z^{*}}{d+1}\big(1+\frac{ a^{d+1}}{2}\big).
\end{eqnarray}
 We can  write the above equation  in the following equivalent form,
\begin{eqnarray}
&&l\approx 2z^{*}b(d,z^{*},z_0),\\&& \nonumber b(d,z^{*},z_0)=\frac{1}{2(d+1)}\big(1+\frac{ a^{d+1}}{2}\big).
\end{eqnarray}
 
Now we can  use the same technique to calculate the area functional. So, we can 
change the variable from $z$ to $\xi$,  use the 
series expansion (which  is valid for all $\frac{\epsilon}{z^{*}}<\xi\ll1$), and  obtain,
\begin{eqnarray}
&&\mathcal{A}(\gamma_A)\approx \frac{L^{d-1} R^d}{(z^{*})^d}\Big(-\frac{1}{d}
(1-(\frac{\epsilon}{z^{*}})^{-d})
+\frac{a^{d+1}}{2}(1-(\frac{\epsilon}{z^{*}})^{d+1})\Big).
\end{eqnarray}
Thus, we obtain both the finite and the divergent parts for this area as, 
\begin{eqnarray}
&&
\mathcal{A}(\gamma_A)\approx  \frac{L^{d-1} R^d}{(z^{*})^d}\Big(-\frac{1}{d}+\frac{a^{d+1}}{2}\Big)+ 
\frac{L^{d-1} R^d}{(z^{*})^d}\frac{1}{d}(\frac{\epsilon}{z^{*}})^{-d}.
\end{eqnarray}
The area used to calculate the 
holographic entanglement entropy is regularized  by subtracting  the 
    the  background AdS geometry from  the   deformed AdS geometry. 
Thus, we use this regularization to remove  the divergent part, and obtain  
the final expression for regularized $\mathcal{A}(\gamma_A)$  as 
\begin{eqnarray}
&&\Delta\mathcal{A}(\gamma_A)\approx  \frac{L^{d-1} R^d}{(z^{*})^d}\Big(-\frac{1}{d}+\frac{a^{d+1}}{2}\Big)
= \frac{L^{d-1} R^d}{(z^{*})^d}\alpha(d,z^{*},z_0),\\&& \nonumber
\alpha(d,z^{*},z_0)=\Big(-\frac{1}{d}+\frac{a^{d+1}}{2}\Big).
\end{eqnarray} 
 Note that from  $\alpha(d,z^{*},z_0)$, we can observe that $z^{*}\neq z_0$.

We can also calculate volume by using the use the same technique. So, we can again  
    change the  variable from $z$ to $\xi$, and use the series expansion, to obtain 
\begin{eqnarray}
&&x_1(\xi)\approx z^{*}\Big(\frac{\xi^{d+2}}{d+2}+\frac{a^{d+1}\xi^{2d+3}}{2d+3}\Big). 
\end{eqnarray}
{Thus, using the series expansions (after integration),  we obtain
\begin{eqnarray}
&&
V(\gamma_A)\approx  \frac{L^{d-1} R^d}{2(d+2)(z^{*})^{d-2}}\Big(1-(\frac{\epsilon}{z^{*}})^2\Big).
\end{eqnarray}
The volume in the deformed AdS geometry is again regularized by subtracting the background AdS geometry from it. 
So, the  regularized volume can be written as
\begin{eqnarray}
&&
\Delta V(\gamma_A)\approx  \frac{L^{d-1} R^d}{(z^{*})^{d-1}}\nu(d,z^{*}),\\&&\nonumber \nu(d,z^{*})=\frac{z^{*}}{2(d+2)}. 
\end{eqnarray}
So, using these results along with the expression for the holographic entanglement entropy and holographic complexity, we obtain 
\begin{eqnarray}
&&\frac{C_A}{S_A} =\frac{n_d l}{R}, 
\end{eqnarray}
 where the regularized value of  the constant  $n_d$ is given by 
\begin{eqnarray}
&&n_d=\frac{\nu(d,z^{*})}{4\pi \alpha(d,z^{*},z_0)b(d,z^{*})}.
\end{eqnarray}
 It may be noted that the effective 
temperature (entanglement temperature) $T_{ent}$ is proportional to the inverse of length of the region, 
as  $T_{ent}=c l^{-1}$ \cite{Bhattacharya:2012mi}. 
So,  we obtain the following universal relation between the holographic entanglement entropy and holographic complexity
\begin{eqnarray}
&&\frac{T_{ent}\mathcal{C}_{A}}{S_A}=\frac{n_d c}{ R}\label{universal}.
\end{eqnarray}


It may be noted that  this relation  is only valid for small subregions.  
This is because  equation used in deriving this relation is only  valid, when  $z<<z_0$
\cite{Bhattacharya:2012mi}.     So,  all the    calculations
have been done for such  a small region,  in which thermal equilibrium is required.  It 
is a region near  the AdS boundary, and thus  the proposed universality conjecture is valid  
 for such a region. 
We would also like to point out that the   metric function $h(z)$ is constructed 
from bulk, and we do  consider any  quantum correction from a  backreaction  on the metric background function.

This is a universal relation between the holographic complexity and  holographic entanglement entropy. 
It exists due to an AdS version of the holographic cavalieri principle, and so we can call this universal relation as the 
  Holographic Cavalieri Principle Conjuncture, and explicitly  state it as following: 
\emph{ Let us assume that     two
regions exist between two parallel AdS slice, and these  two regions are  codimension two slice of an asymptotically AdS space, such that they
have equal areas, and the    CFT  duals to these 
two regions have equal entangled temperature, then they will have  equal holographic complexity.}

This conjecture can have a lot of applications as it can also be used to obtain a definition of complexity for 
the boundary theory. There is no agreed definition of the complexity for the boundary theory, however, using this conjecture, we can define 
the complexity of the boundary theory. Thus, we can obtain the complexity for the boundary theory using 
$ \mathcal{C}_{A} = {c\cdot n_d}S_A/{R}T_{ent} $, because 
the boundary dual of all 
these quantities except complexity is well defined. So, this relation can be used as a definition for the complexity 
of the boundary theory. It is also expected that the holographic complexity would be directly proportional to the holographic 
entanglement entropy, as the difficulty  of obtaining the information from a system will increase as the amount of information 
lost from a system will increase. Thus, this result is something we would expect on physical grounds for the boundary theory.

 \section{Circular Disk } 
 In this section, we will demonstrate that this conjecture holds for a circular disk.  
The holographic entanglement entropy \cite{RT} and holographic complexity  \cite{Alishahiha:2015rta} for such a geometry has already been analyzed.  
We would like to point out that due to our definition of the area and volume of the minimal surface, we will only use the finite part of 
the holographic entanglement entropy and holographic complexity. This is because these quantities are regularized by subtracting the contributions 
coming from the background AdS. Now using the   bulk of  AdS$_{d+2}$, it is possible to  define a 
  sphere of radius $l$. 
So, we  parametrize such a  metric as  
\begin{eqnarray}
&&ds^2=\frac{R^2}{r^2}\Big(-dt^2+dr^2+d\rho^2+\rho^2 d\Omega_{d-1}\Big) \label{metric}.
\end{eqnarray}
The entangling region is represented by $\{t=0,r\leq l\}$,  where $l$ is the radius of a circular disk. 
The area  and volume functionals  for the  parametrization $\rho=\rho(r)$, can be written as 
\begin{eqnarray}
&& \mathcal{A} (\gamma_A)=\Omega_{d-1}R^d\int_{\epsilon}^{l}dr \frac{\rho(r)^{d-1}}{r^d}\sqrt{1+(\frac{d\rho(r)}{dr})^2},\label{A11}
\\&&
V(\gamma_A)=\frac{\Omega_{d-1}R^{d+1}}{d}\int_{\epsilon}^{l}dr \frac{\rho(r)^d}{r^{d+1}}.\label{V11}
\end{eqnarray}
Now we can write the solution of equation of motion describing this system as   $\rho(r)=\sqrt{l^2-r^2}$. 
Using this solution, we can obtain an expression for Eqs. (\ref{A11},\ref{V11}),  
\begin{eqnarray}
&& \mathcal{A}(\gamma_A)=\Omega_{d-1}R^d \int_{\epsilon}^{l}dr \frac{(l^2-r^2)^{d/2-1}}{r^d},
\\&&
V(\gamma_A)=\frac{\Omega_{d-1}R^{d+1}}{d}\int_{\epsilon}^{l}dr \frac{(l^2-r^2)^{d/2}}{r^{d+1}}.
\end{eqnarray}
Now we expand the above integrals in series,and use the suitable regularization for them. So, we are only left with the finite part of the 
 volume and area functionals,
\begin{eqnarray}
&& \mathcal{A}(\gamma_A)=\Omega_{d-1}R^d\sum_{n=0}^{\infty}\frac{(1-d/2)_n}{n!(2n-d+1)},
\\&&
V(\gamma_A)=\frac{\Omega_{d-1} R^{d} l}{d}\sum_{n=0}^{\infty}\frac{(-d/2)_n}{n!(2n-d)}.
\end{eqnarray}
Now using (\ref{HEE},\ref{HC}),  we obtain
\begin{eqnarray}
&&S_A=\frac{\Omega_{d-1}R^d}{4G}\sum_{n=0}^{\infty}\frac{(1-d/2)_n}{n!(2n-d+1)}
\\&&
\mathcal{C}_A= \frac{\Omega_{d-1} R^d l}{8\pi R G d}\sum_{n=0}^{\infty}\frac{(-d/2)_n}{n!(2n-d)}
\end{eqnarray}
So, we can write the ratio of such terms as   
\begin{eqnarray}
&&\frac{\mathcal{C}_A}{S_A}=\frac{c_d l}{R}\label{pure}.
\end{eqnarray}
Thus, we observe that even for this geometry, the holographic entanglement entropy is propotional to the holographic complexity. 
This is the expression we expected from our conjecture. Thus, this conjecture holds for the such a geometry.  

\section{ Asymptotically AdS Spacetime}
We can try to argue that such a conjecture is justified for a general asymptotically AdS$_{d+1}$.
The appropriate form of the metric written in Fefferman-Graham coordinates is given by 
\begin{eqnarray}
&&ds^2_{d+1}=\frac{R^2}{r^2}\Big(dr^2+g_{\mu\nu}dx^{\mu}dx^{\nu}\Big)
\label{g3}
\end{eqnarray}
We choose an entangled strip parametrized by $\{t=0,x_1=x(r)\in[-l/2,l/2],x_i\in[0,L],i=2,..,d-1$, and write the  area functional for this metric as 
\begin{eqnarray}
&&\mathcal{A}(\gamma _{A})=R^{d-1}\int^{d-2}xdr\frac{\sqrt{g(r)(1+G(r)x'^2)}}{r^{d-1}}
\label{Area3}
\end{eqnarray}
where $g\equiv |g_{ij}|$,$G(r)=g_{11}-\frac{g_{1i}g_{j1}}{g_{ij}}$. A  conserved charge can be constructed using this area functional  (for general $x(r)$)
\begin{eqnarray}
&&(\frac{R}{r})^{d-1}\frac{\sqrt{g(r)}G(r)x'}{\sqrt{g(r)(1+G(r)x'^2)}}=(\frac{R}{r^{*}})^{d-1}\sqrt{g(r^{*})G(r^{*})}
\label{xprime}
\end{eqnarray}
Thus, we obtain the following result, 
\begin{eqnarray}
&&x(r)=\int dr \frac{(\frac{R}{r^{*}})^{d-1}\sqrt{g(r^{*})G(r^{*})}}{\sqrt{G(r)\Big(g(r)G(r)(\frac{R}{r})^{2(d-1)}-g(r^{*})G(r^{*})(\frac{R}{r^{*}})^{2(d-1)}
\Big)}}
\end{eqnarray}
Total entangled length $l$ and the total volume $ V(\gamma_A)$,  can be written as 
\begin{eqnarray}
 l&=&2\int_{0}^{r^{*}}\frac{(\frac{R}{r^{*}})^{d-1}\sqrt{g(r^{*})G(r^{*})}}{\sqrt{G(r)\Big(g(r)G(r)(\frac{R}{r})^{2(d-1)}-g(r^{*})G(r^{*})(\frac{R}{r^{*}})^{2(d-1)}
\Big)}}. \\ 
  V(\gamma_A)&=&\int d^{d-2}x \int_{0}^{r^{*}}x(r) (\frac{R}{r})^{d}\sqrt{|g_{ij}|}\sqrt{1+\frac{g_{1i}g_{j1}}{g_{ij}}}dr
\end{eqnarray}
As it is not possible to explicitly calculate the holographic entanglement entropy for a general metric, we will simplify our analysis to 
the general form of the metric on the AdS$_3$.  
The metric of any asymptotically AdS$_3$   can be represented by Eq. (\ref{g3}) 
when $g_{\mu\nu}=h_{\mu\nu}r^d$, where $h_{\mu\nu}$ is a uniform metric. Thus, for such a metric,  we can easily integrate Eq. (\ref{xprime}),  and obtain 
\begin{eqnarray}
&&x(r)=\frac{1}{\sqrt{H}\cosh(\frac{r}{r^{*}})}
\end{eqnarray}
where $H=h_{11}-\frac{h_{1i}h_{j1}}{h_{ij}}$. Now for AdS$_3$, we can obtain  the area as 
$ \mbox{Area}=2 R \sqrt{|h_{\mu\nu}|}$, and so the 
entanglement entropy can be written as 
\begin{eqnarray}
&& S_A=\frac{R \sqrt{|h_{\mu\nu}|}}{2G_3}
\end{eqnarray}
Similarly, we can  calculation the volume, and express it as 
$ V(\gamma_A)=R^2\sqrt{|h_{\mu\nu}|} \mathcal{N}
$
where $\mathcal{N}=\int_{\epsilon}^{1}\frac{d\xi}{\xi \cosh(\xi)}\approx -\frac{19}{96}-\ln(\epsilon)+\mathcal{O}(\epsilon^6)\gg 1
$, and so the 
holographic complexity can be written as 
\begin{eqnarray}
&&\mathcal{C}_{A}=\frac{R\sqrt{|h_{\mu\nu}|} \mathcal{N}}{8\pi G_3}. 
\end{eqnarray}
Furthermore, the total length is given by
\begin{eqnarray}
&&l=\frac{2}{\sqrt{H}}\mathcal{B}
\end{eqnarray}
where $\mathcal{B}=\int_{0}^{1}\frac{d\xi}{\xi\sqrt{\xi^2-1}}=-\frac{\pi}{2}-i\ln\frac{\epsilon}{2}+\mathcal{O}(\epsilon^6)\gg 1$. Note that $\sqrt{H}\sim \frac{1}{\sqrt{|h_{ij}|}}$. 
If we combine three equations and using the definition of the temperature,  we obtain 
\begin{eqnarray}
&&\frac{T_{ent}\mathcal{C}_{A}}{S_A}=\frac{\mathcal{N}}{8\pi \mathcal{B} R}\label{universal3}.
\end{eqnarray}
Even though we have explicitly calculated this for all asymptotically AdS$_3$, we can follow the same algorithm and calculate these quantities for any 
asymptotically AdS metric. However, such calculation, even though conceptually straightforward, can become computational complicated. So, in the next
section, we will demonstrate this conjecture holds for an important asymptotically AdS. 
\section{Janus solution}
Now we will explicitly test this conjuncture for Janus solution. First of all, we will consider a 
  AdS$_3$ Janus solution \cite{r5}.
The  Janus solution   interpolates between two AdS spaces \cite{Bak:2003jk}. 
This bulk model for this solution is  defined by the following action, 
\begin{eqnarray}
S=-\frac{1}{16\pi G_N}\int dx^3 \sqrt{g}
\left({\cal R} - g^{a b} \partial_a \phi \partial_b \phi +\frac{2}{R^2}\right).
\label{janus-acf}
\end{eqnarray}
 The Janus solution is a three dimensional (actually the simplest analytic) member of the generally $AdS_d$-sliced
 domain walls with the isometry group $SO(d-1,2)$  \cite{l1, l2, Sabra, sa, as, Behrndt}. The  metric for such a solution can be 
 written as 
 \begin{eqnarray} \label{adswall}
ds^2 = e^{2A(r)} g_{ij}(x) dx^i dx^j + e^{2h(r)} dr^2,
\end{eqnarray}
where $g_{ij}(x)$ is a metric on $AdS_d$ with scale $R_d$. It may be noted that such a 
domain wall has also been obtained   as a
solution in the  type IIB supergravity  \cite{Bak:2003jk}. This solution 
has no    $r$-dependent matter fields except a flowing dilaton $\phi(r)$. 
This solution is
regular, if parameters are chosen  such that the rate of variation
of the dilaton is sufficiently slow. We  use the radial coordinate
$r$ for which $h(r) = 0$,  so we take $\phi' =c \exp(-dA(r))$, and by this the  scalar
equation of  is also satisfied. Now the wall profile
equation can be written as 
 \begin{eqnarray} \label{profeq}
A'^2 = (1/L^2) [1 -e^{-2A} + b e^{-2dA}],
\end{eqnarray}
The constant $b$ is related to $c$ by $b=\frac{\kappa^2}{d(d-1)} c^2R^2$.
We have set $R_d=L$ for simplicity. So, 
when $b=0$, the solution gives pure AdS$_{d+1}$, 
 \begin{eqnarray}
ds^2 = \cosh^2(r/L) g_{ij}(x) dx^i dx^j  +  dr^2.
\end{eqnarray}
However, for $b \ne 0$,  it is not possible to obtain a simple solution for Eq.   (\ref{profeq}), 
 (unless $d=2$ and this is the Janus metric). The metric for $AdS_2$ sliced domain walls in $AdS_3$  
 can be written as 
 \begin{eqnarray}
ds^2 = e^{2A(r)} ds_{AdS_2}^2 + dr^2.
\end{eqnarray}
The explicit solution of the equations of motion is the Janus solution, is given by the metric
\begin{eqnarray}
&& ds^2=R^2\left(dy^2+ f(y) ds^2_{AdS2}\right), \\
&& f (y) = \frac{1}{2} (1+\sqrt{1-2\gamma^2} \cosh(2y) ),
\label{eq: janus-metric}
\end{eqnarray}
and the dilaton is given by the function
\begin{eqnarray}
\phi (y) = \gamma\int^{y}_{-\infty}\frac{dy}{f (y)}+\phi_1, \label{phj}
\end{eqnarray}
where $\gamma~ (\leq \frac{1}{\sqrt{2}})$ is the parameter of Janus deformation.

The metric of AdS$_2$ slice is given by 
\begin{equation}
ds^2_{AdS2}=(dz^2+dx^2)/z^2,\ \ \phi_1=\phi(-\infty), 
\end{equation}
 and it  is dual
to the coupling constant of the exactly marginal deformation
for the ground state $|\Omega_1>$. The value $\phi_2=\phi(\infty)$ for the other
ground state $|\Omega_2>$ is obtained by performing the integral give in Eq. (\ref{phj}). So, we have 
\begin{equation}
 \phi_2-\phi_1=\sqrt{2}\arctan\left[\frac{1-\sqrt{1-2\gamma^2}}{\sqrt{2}\gamma}\right]\simeq \gamma, 
\end{equation}
when $\gamma\ll 1$.
Now if  the bulk extension
of the surface is parameterized by $x=x(r)$, then the  corresponding area is given by
\begin{eqnarray}
&&\mathcal{A}(\gamma _{A})=\int\frac{R^2 f(0)}{z^2}\sqrt{1+x'(z)^2}dz. \label{area}
\end{eqnarray}
The minimal surface $x(z)$ is the solution of the following equation, 
\begin{eqnarray}
&&\frac{x'(z)}{z^2\sqrt{1+x'(z)^2}}=\frac{1}{(z^{*})^2}, \label{first}
\end{eqnarray}
 with the auxiliary boundary condition,  $x'(z)|_{z=z^{*}}=\infty$. 
Now the  solution of this equation  can be written as
\begin{eqnarray}
&&x(z)=z^{*}\Big(E(\xi,i)-F(\xi,i)\Big)\label{sol}
\end{eqnarray}
where $\xi=\frac{z}{z^{*}}$, and $E(z,k),F(z,k)$ are elliptic functions,  
\begin{eqnarray}
&&E(z,k)=\int_{0}^z\frac{\sqrt{1-k^2r^2}}{\sqrt{1-r^2}}dr,\\
&&F(z,k)=\int_{0}^z\frac{dr}{\sqrt{1-k^2r^2}\sqrt{1-r^2}}. 
\end{eqnarray}
Thus,   minimizing the area, we obtain 
\begin{eqnarray}
&&\mathcal{A}(\gamma _{A})=\frac{R^2 f(0)}{z^{*}}\int_{\frac{\epsilon}{z^{*}}}^1\frac{d\xi}{\xi^2\sqrt{1-\xi^4}}
\end{eqnarray}
where $z^{*}$ is turning point and $\epsilon$ is a UV cut off. 
Now for the solution Eq. (\ref{sol}), the  total entangled length $l$, the finite part of the entanglement 
 area functional $A(\gamma)$ and volume of codimension one time slice $V(\gamma_A)$ of the metric are obtained as follows, 
\begin{eqnarray}
&&l=2z^{*}\Big(E(1,i)-F(1,i)\Big)
 \label{l3}
\\&& \mathcal{A}(\gamma _{A})=\frac{R^2 f(0)}{z^{*}}(-1+\sum_{n=1}^{\infty}\frac{(1/2)_n}{n!(4n-1)})
\label{A3}
\\&&
V(\gamma_A)=R^2\big(\frac{\pi}{4}+K(i)-E(1,i)
\big)
\label{V3}.
\end{eqnarray}
Here $K(k)$ is another  elliptic functions,
\begin{eqnarray}
&&K(k)=\int_{0}^1\frac{dr}{\sqrt{1-k^2r^2}\sqrt{1-r^2}}
\end{eqnarray}
 So, using Eqs. (\ref{l3}-\ref{V3}), it is possible to explicitly demonstrate that    the holographic Cavalieri principle holds for this solution, 
\begin{eqnarray}
&&\frac{T_{ent}\mathcal{C}_{A}}{S_{A}}=\frac{\mathcal{N}}{R}\label{universal2}.
\end{eqnarray}
where $\mathcal{N}=  \frac{c\cdot n_1}{\pi n_2 f(0)}$ is a numeric factor, 
and $n_1, n_2$ are given by 
\begin{eqnarray}
n_1&=& \frac{\frac{\pi}{4}+K(i)-E(1,i)}{2\Big(E(1,i)-F(1,i)\Big)},
\nonumber \\ n_2&=&-1+\sum_{n=1}^{\infty}\frac{(1/2)_n}{n!(4n-1)}.
\end{eqnarray}
Thus, we have been able to explicitly demonstrate that holographic Cavalieri principle hold for Janus solution. 
This can be used to analyse the holographic complexity for the boundary theory dual to the Janus solution. 
It may be noted that Janus solution is dual to an interesting field theoretical system. 
This is because the  field theoretical system  dual to Janus solution is a boundary spacetime divided by 
a codimension one defect \cite{Bak:2003jk}.  A  different Yang-Mills coupling exists in each of the two halves of this boundary 
spacetime. In fact, the  
      the string theoretical configurations    for this solution have also been analysed \cite{Bak:2003jk}. 
  The    conformal perturbation theory has been used to analyse the       
     quantum level conformal symmetry of the Janus solution 
 \cite{j1}. 
The holographic entanglement entropy  for Janus solution  has been calculated,  and it has been used 
for analyzing the behavior of boundary  theory dual to the Janus solution \cite{entr}. 
The holographic complexity can also be used to analyse the behavior of the boundary theory dual to the Janus solution. 
So, it  would   be interesting   to use the results of this paper to analyse the behavior of the boundary  theory 
dual to Janus solution. 
\section{Application}
In this paper, we have used a proposal for holographic complexity, 
which states that the holographic complexity of a system is equal to the volume enclosed by  a minimal surface. 
There is another recent proposal for the holographic complexity and this proposal states that the holographic complexity of a system 
is equal to the bulk action, calculated on a  Wheeler-DeWitt patch \cite{ Brown:2015bva, brown} 
\begin{eqnarray}  
 \mathcal{C}= \frac{A}{\pi \hbar}, \label{ref}
\end{eqnarray}
where $A$ is the   bulk action evaluated on
the Wheeler-deWitt patch with a suitable  temporal boundary, and $\mathcal{C} $ is this holographic   complexity obtained using this new proposal. 
It is possible to calculate the action on a Wheeler-DeWitt patch for such geometries, 
using the null boundaries of the Wheeler-DeWitt patch.  

It has been argued that black holes saturate the bound for rate of change of complexity, and so for black holes the rate of change of complexity 
is given by \cite{Brown:2015bva, Lloyd}
\begin{eqnarray}
&&\frac{d\mathcal{C} }{dt} =  {2M},
\end{eqnarray}
where $M$ is the mass of the black hole. 
Thus, it was argued that the black holes are the fastest computes. 

We will demonstrate that this result can also be obtained from the other   proposals for 
holographic complexity (in which holographic complexity is related to the volume),
using the  conjuncture presented in this paper. Thus, we will argue   that
both these proposals for holographic complexity  can represent the same physics.
However, here we will use the holographic complexity for a subregion, and we so we need to first 
define the growth for it. It has been demonstrated it is possible to define the growth of such a volume 
 \cite{bh17, bh14}, and so we can analyze such a growth of holographic complexity for such a subregion,  in this paper.
 
It is  known that the holographic entanglement 
entropy scales as $L^2$ (because of the definition of the area used to calculate it). Furthermore, the entanglement 
temperature can be represented as 
$T_{ent}\sim L^{-1}$, 
where $L$ the size of entangled region. So,  
$T_{ent}\mathcal{C}_A$ represents the  time variation of $\mathcal{C}_A$.  
Now using the conjecture proposed in this paper, we obtain 
\begin{eqnarray}&&
\frac{d\mathcal{C}_A}{dt }\approx\frac{S_A}{R}\sim L\sim \mbox{size of horizon}\leq 2M
\end{eqnarray}
 where $t \approx T_{ent}^{-1}$, and we have also assumed that the
 entangled region size  remains in the limit of $L\leq \mbox{size of horizon}$. 
 Thus, we would obtain such a time from the entanglement temperature of the dual conformal field theory. 
So, in the limiting case for  black holes with mass $M$, we obtain 
\begin{eqnarray}\label{Lloyd}
&&\frac{d\mathcal{C}_A}{dt}= {2M} .
\end{eqnarray}
We again conclude that the black holes are fastest computers, however, we have now obtained this result using the proposal that the 
holographic complexity equals to the  volume. 
Thus, we have demonstrated that the complexity defined in Eq. (\ref{HC}) along with Eq. (\ref{universal}), produces the same physics as 
as the complexity defined in Eq. (\ref{ref}). 
So,   the proposal that holographic complexity is equal to the action, 
and the proposal that the holographic complexity is equal to the volume represent the same physics.

\section{Conclusion}
In this letter, we propose that a non-trivial but universal relation exists between the holographic
quantum complexity   and the holographic entanglement entropy. 
As this relates a quantity which is dual to a volume in AdS to a quantity which is dual to an area in AdS, it can 
be considered as a  holographic version of Cavalieri principle. Furthermore, in analogy with  the usual   Cavalieri principle, the regions 
analysed were assumed to  exist between two parallel AdS slice. We argued that such a conjuncture should hold in general, as it is based on 
the AdS version of the cavalieri principle.  We also explicitly demonstrated this to be the case for AdS$_3$. However, as it is not 
possible to obtain a general expression for the holographic entanglement entropy, we made a conjecture that such a universal relation should hold. 
This is because the higher dimensional case would be conceptually similar to this case, however, they would be computationally more complicated. 
We demonstrated that this conjecture holds for a circular disk. 
We also  explicitly demonstrate that our conjuncture hold for Janus solution, which has recently 
been obtained in type IIB string theory. 

We also analyzed the relation between the proposal that the holographic complexity equal to the action and the holographic complexity 
equal to the volume. It was already demonstrated that the black holes are the fastest computers using the proposal which stated that the 
holographic complexity is equal to the action evaluated at the Wheeler-DeWitt patch \cite{ Brown:2015bva, brown}. In this paper, we demonstrated 
that the black holes are the fastest computers using the proposals which states that the holographic complexity is equal to the volume. 
This has done by using the universal relation between the holographic complexity and holographic entanglement entropy proposed in this paper. 
Thus, we demonstrated that both these proposals represent the same physics. 

It is possible to apply this conjuncture to  various different excited AdS solutions. So, 
it would be interesting to analyse the consequences of this conjuncture, and use it for 
analyse the behavior of holographic complexity and holographic entanglement entropy for various different excited AdS states. 
We also proposed that this universal relation can be used to obtain a definition of complexity for the boundary conformal field theory. 
It would be interesting to analyse the implications of this relation for the fidelity susceptibility  \cite{r5}, 
as fidelity susceptibility  can be used to analyse the quantum phase transitions in field theories dual to different deformed AdS solutions \cite{r6,r7, r8}.
It might be possible to obtain a relation between the fidelity susceptibility and holographic complexity, and then use this relation along with 
the results obtained in this paper to analyse the quantum phase transitions in different  systems. 
It will also be interesting to analyse the consequences of this conjuncture on the black hole information paradox, as both holographic complexity
has become more important than holographic entanglement entropy as it has been argued that information might not be ideally lost, but it might
be impossible to recover this information from Hawking radiation \cite{hawk}. So, it seems holographic complexity might be used for  understanding the black hole 
information paradox, and the results of this paper can be used for analysing the behavior of holographic complexity. 

It may be noted that there  are other   ways to 
define this volume in the bulk, so various different proposals have been used for holographic complexity.
The complexity can also be 
calculated from the  maximal volume in the AdS which
ends on the time slice at the AdS boundary \cite{r5}.  It has been demonstrated that this proposal corresponds to the 
fidelity susceptibility   of the boundary conformal field theory, 
and so it is important for   analyzing   the quantum phase transitions \cite{r6,r7, r8}.
However, in this paper, we have  not use such proposals, and   restrict our analysis to the volume  
 enclosed by the    minimal surface   used for calculating the holographic 
entanglement entropy \cite{Alishahiha:2015rta}. It would be interesting to analyze a relation between the holographic complexity, entanglement 
entropy and fidelity susceptibility. It may be noted that such a relation exists for a D3-brane geometry \cite{brane}, but it will be interesting 
to investigate if such a relation exists for other deformed geometries. It would also be interesting to investigate this for time dependent geometries, 
as holographic complexity \cite{time1} and holographic entanglement entropy \cite{time2} have been studied for such geometries. 


\end{document}